\def\asca{{\it ASCA\/}}
\def\chandra{{\it Chandra\/}}

\def\rosat{{\it ROSAT\/}}

\def\ltsima{$\; \buildrel < \over \sim \;$}
\def\simlt{\lower.5ex\hbox{\ltsima}}
\def\gtsima{$\; \buildrel > \over \sim \;$}
\def\simgt{\lower.5ex\hbox{\gtsima}}

\documentclass[preprint]{aastex}

\begin{document}


\title{X-ray Sources in the Hubble Deep Field Detected by Chandra}


\author{A.E.~Hornschemeier,\footnotemark\ ~
W.N.~Brandt,$^1$
G.P.~Garmire,$^1$
D.P.~Schneider,$^1$
P.S.~Broos,$^1$
L.K.~Townsley,$^1$
M.W.~Bautz,\footnotemark\ ~
D.N.~Burrows,$^1$
G.~Chartas,$^1$
E.D.~Feigelson,$^1$ 
R.~Griffiths,\footnotemark\ ~
D.~Lumb,\footnotemark\ ~
J.A.~Nousek$^1$ and
W.L.W.~Sargent\footnotemark\ 
}

\footnotetext[1]{Department of Astronomy \& Astrophysics, 525 Davey Laboratory, 
Pennsylvania State University, University Park, PA 16802}

\footnotetext[2]{Massachusetts Institute of Technology, Center for Space Research, 
70 Vassar Street, Building 37, Cambridge, MA 02139}

\footnotetext[3]{Department of Physics, Carnegie Mellon University, Pittsburgh, PA 15213}

\footnotetext[4]{Astrophysics Division, ESTEC, Keplerlaan~1, 2200~AG Noordwijk, The Netherlands}

\footnotetext[5]{Palomar Observatory, California Institute of Technology, Pasadena, CA 91125}


\begin{abstract}
We present first results from an X-ray study of the Hubble Deep Field North (HDF-N) 
and its environs obtained using 166~ks of data collected by the Advanced CCD 
Imaging Spectrometer (ACIS) on board the \chandra\ X-ray Observatory. 
This is the deepest X-ray observation ever reported, and in the HDF-N 
itself we detect six X-ray sources down to a 0.5--8~keV flux limit of 
$\approx 4\times 10^{-16}$~erg~cm$^{-2}$~s$^{-1}$. 
Comparing these sources with objects seen in multiwavelength HDF-N studies
shows positional coincidences with
the extremely red object NICMOS~J123651.74 +621221.4, 
an active galactic nucleus (AGN), 
three elliptical galaxies, and
one nearby spiral galaxy. 
The X-ray emission from the ellipticals is consistent with that
expected from a hot interstellar medium, and the spiral galaxy
emission may arise from a `super-Eddington' X-ray binary or 
ultraluminous supernova remnant. 
Four of the X-ray sources have been detected at radio wavelengths. 
We also place X-ray upper limits on AGN candidates found in the HDF-N, 
and we present the tightest constraints yet on X-ray emission 
from the SCUBA submillimeter source population. None of the 10
high-significance submillimeter sources reported in the HDF-N and
its vicinity is detected with \chandra\ ACIS. These sources appear 
to be dominated by star formation or have AGN with Compton-thick tori 
and little circumnuclear X-ray scattering. 
\end{abstract}




\section{Introduction}

The Hubble Deep Field North (HDF-N; Williams et~al. 1996, hereafter W96) 
provides an unprecedented view of the distant 
Universe, and the incredible amount of work focused on 
this patch of sky has led to fundamental insights about the formation
and evolution of galaxies. Intense follow-up studies have been
performed at a variety of wavelengths including the radio, 
submillimeter, far-infrared, near-infrared, and optical bands; 
these investigations have discovered a fascinating array of cosmic 
objects. Notably lacking until now, however, is a deep X-ray survey 
of this area.\footnote{The only previous pointed X-ray observation of 
the HDF-N was a 21~ks exposure made with the \rosat\ High Resolution 
Imager (HRI). No X-ray sources in the HDF-N were detected due to 
the limited sensitivity of this observation, and to our knowledge
the \rosat\ results have never been published.}
X-ray surveys provide the most direct and unbiased probe of massive 
black hole accretion activity throughout the Universe. This is particularly 
true in the hard X-ray band above 2~keV where most of the energy density 
of the extragalactic X-ray background (XRB) resides. In addition, X-ray
surveys allow studies of starburst galaxies, elliptical galaxies, 
clusters and groups of galaxies, and other 
objects. Early observations with the \chandra\ X-ray 
Observatory (hereafter \chandra; Weisskopf, O'Dell \& van~Speybroeck 1996) 
have resolved much ($\simgt$~60--75\%) of the 2--8~keV XRB into point 
sources (Brandt et~al. 2000; Mushotzky et~al. 2000); an even higher
fraction ($\simgt$~70--80\%) of the 0.5--2~keV X-ray 
background had already been resolved by \rosat\
(e.g., Hasinger et~al. 1998). The key goals now remaining are to 
resolve the rest of the XRB and to understand in detail the nature 
and evolution of the sources creating it. 

In this letter, we present first results from a deep X-ray survey of
the HDF-N area using the \chandra\ Advanced CCD Imaging Spectrometer 
(ACIS; G.P. Garmire et~al. 2000, in preparation). We focus on X-ray 
sources in the HDF-N itself and in its immediate vicinity where
substantial follow-up work has already been performed. 
The Galactic column density along this line of sight is
$(1.6\pm 0.4)\times 10^{20}$~cm$^{-2}$ (Stark et~al. 1992),
and $H_0=70$~km~s$^{-1}$ Mpc$^{-1}$ and $q_0=0.5$
are adopted. 
			

\section{Chandra ACIS Observations and Data Analysis}

The field containing the HDF-N was observed with the \chandra\ ACIS
for a total exposure time of 166~ks in three segments on 
1999~November~13 (50~ks),
1999~November~14 (58~ks), and 
1999~November~21 (58~ks). 
The HDF-N was placed near the aim point for the ACIS-I 
array on CCD~I3 during all three observations. 
The full ACIS-I field of view is
$16^{\prime}\times 16^{\prime}$, and the
on-axis image quality is $\approx 0.5^{\prime\prime}$ FWHM. 
The observations were corrected 
for the radiation damage the CCDs suffered 
during the first few weeks of the mission 
(Prigozhin et~al. 2000; Hill et~al. 2000)
following the procedure of Townsley et~al. (2000), which corrects
simultaneously for both position-dependent gain shifts and
event grade changes. The three exposures were then co-added
after cross-registration using bright sources near the optical 
axis; we believe cross-registration to be accurate to within 
$0.3^{\prime\prime}$. Absolute X-ray source positions are
accurate to within $1^{\prime\prime}$ (see \S3.1). 


We created images from 
0.5--8.0~keV (full band), 
0.5--2.0~keV (soft band) and 
2--8~keV (hard band) for further study, and we use
\asca\ event grades 0, 2, 3, 4 and 6 in all analyses.
We searched these images with the {\sc wavdetect} 
software (Dobrzycki et~al. 1999; Freeman et~al. 2000) 
using the same methods and safety checks as
Brandt et~al. (2000). We have used a probability 
threshold of $1\times 10^{-7}$, and light curve 
analyses verify that none of the X-ray sources discussed
below is affected by a `flaring pixel' problem
(\chandra\ X-ray Center, private communication). 
All of the sources below are consistent with being pointlike, 
although the constraints are not tight in most cases due to 
limited numbers of photons. 
Even with a 166~ks observation the data are far from being background 
limited (the full-band background is 0.037~count~pixel$^{-1}$), 
and within $\approx 3^\prime$ of the aim point the detection limit 
with our selection criteria is $\approx 7$ photons 
in the full, soft and hard bands. For 
a power-law model with photon index $\Gamma=2$ 
and the Galactic column density, this corresponds to a 
soft (hard) observed flux detection limit of 
$2.3\times 10^{-16}$~erg~cm$^{-2}$~s$^{-1}$
($8.8\times 10^{-16}$~erg~cm$^{-2}$~s$^{-1}$). 
For the cosmology of \S 1, at $z=1$ the corresponding rest-frame luminosity limits are
$L_{1-4~{\rm keV}}=7.3\times 10^{41}$~erg~s$^{-1}$
($L_{4-16~{\rm keV}}=2.6\times 10^{42}$~erg~s$^{-1}$); even
fairly low-luminosity Seyfert galaxies should be
detected at this redshift.  For an open Universe with $q_0=0.1$, these 
limits become $L_{1-4~{\rm keV}}=1.0\times 10^{42}$~erg~s$^{-1}$
($L_{4-16~{\rm keV}}=3.7\times 10^{42}$~erg~s$^{-1}$).


\section{Results for Sources in and near the HDF-N}

\subsection{The X-ray Sources in the HDF-N}

We have detected six full-band X-ray sources in the HDF-N 
(see Table~1 and Figure~1). This number of detections 
is consistent with plausible extrapolations of the number counts. 

Four of the X-ray detections are positionally coincident 
(to within $0.8^{\prime\prime}$) with 
8.5~GHz sources from Richards et~al. (1998, hereafter R98), giving us 
confidence in these matches and, more generally, in the X-ray positions.
{\bf CXOHDFN~J123646.4+621404} ($z=0.960$) is a red spiral galaxy oriented 
nearly face-on that hosts an active galactic nucleus 
(AGN; Phillips et~al. 1997). Variable radio and optical emission have 
been detected (R98; Sarajedini et~al. 2000). 
{\bf CXOHDFN~J123651.8+621220} is spatially coincident with the remarkable source 
NICMOS~J123651.74+621221.4, the second reddest 
object in the NICMOS survey of Dickinson et al. (2000). This source is also
an 8.5~GHz emitter, and it may have 15~$\mu$m and 1.3~mm emission as well
(R98; Aussel et~al. 1999; Downes et~al. 1999). While its continuum is very 
red, it is detected in $B_{450}$, $V_{606}$ and $I_{814}$ 
and is thus unlikely to be an extremely 
high redshift object; photometric redshift estimates indicate
$z=$~2.6--2.7, but these have significant uncertainty due 
to possible reddening by dust (M. Dickinson, private communication). 
CXOHDFN~J123651.8+621220 has the largest hard-band to soft-band 
count ratio (hereafter `band ratio') of any X-ray source
in the HDF-N (see Table~1). Provided the underlying
X-ray continuum shape is typical for an AGN with $\Gamma=$~1.7--2.1, 
the large band ratio suggests substantial internal absorption by 
a column density $N_{\rm H}\simgt 10^{23}$~cm$^{-2}$. If the photometric
redshift is correct the implied X-ray luminosity is 
$\simgt 10^{44}$~erg~s$^{-1}$, and this source would be a 
good candidate for a type~2 Quasi-Stellar Object. 

{\bf CXOHDFN~J123655.5+621310} ($z=0.968$) and {\bf CXOHDFN~J123657.0+621301} ($z=0.474$) 
are both ellipticals with radio emission. The rather flat radio spectrum 
($\alpha_{\rm r}<0.3$) of the first may suggest AGN activity, although 
the X-ray luminosity for this object
is not unusually high given the large $B$-band 
luminosity [$\log(L_{\rm B})=44.52$; compare with Figure~1 of 
Eskridge, Fabbiano \& Kim 1995]. This source has also been detected 
in the Hogg et~al. (2000) survey of the HDF-N at 3.2~$\mu$m. The second
source has a steep radio spectrum ($\alpha_{\rm r}=1.0\pm 0.3$) as well as 
a possible ISO detection, and its X-ray luminosity is consistent with that
expected from hot gas in an elliptical. Neither of these ellipticals
is detected in the hard band, although we cannot rule out the 
presence of hard power-law emission with luminosity comparable to
that seen in some nearby ellipticals (Allen, Di~Matteo \& Fabian 2000). 

The two X-ray sources that are not positionally coincident with radio 
sources are coincident with fairly bright W96 sources. 
{\bf CXOHDFN~J123641.9+621131} appears to be matched with a bright 
($V_{606}=20.0$) foreground spiral galaxy at $z=0.089$ lying near 
the edge of the HDF-N. Somewhat surprisingly, however, the X-ray 
source position is not consistent with the nucleus of this galaxy. 
It may be coincident with a bright spot seen clearly in the 
$U_{300}$ image of W96. 
If the identification with the galaxy 
is correct, the X-ray source is $\sim 3.5$~kpc from the nucleus and
its 0.5--8~keV luminosity is $7.2\times 10^{39}$~erg~s$^{-1}$, low 
enough that this source might plausibly be an off-nuclear 
`super Eddington' X-ray binary or ultraluminous supernova remnant 
(e.g., Fabbiano 1998).  The galaxy may also have some 8.5 GHz
emission, although this emission is not positionally consistent with the X-ray
source (see Table 5 of R98).
{\bf CXOHDFN~J123648.1+621309} ($z=0.476$) is another elliptical that has
been detected at 3.2~$\mu$m by Hogg et~al. (2000); its X-ray
luminosity is consistent with hot gas emission. 

\subsection{Active Galactic Nucleus Candidates in the HDF-N}

X-ray emission is a universal property of AGN. Several 
surveys for AGN in the HDF-N itself have 
been performed, and we have searched for X-ray emission from 
the resulting AGN candidates of Jarvis \& MacAlpine (1998), 
Conti et~al. (1999) and Sarajedini et~al. (2000). Aside from 
CXOHDFN~J123646.4+621404, the $z=0.960$ AGN discussed above, we do
not detect any of these candidates, and we place X-ray upper limits
as discussed at the end of \S2. The physical implications of these
upper limits vary depending upon the properties (e.g., optical
magnitudes and redshifts) of the individual candidates; given the
typical range of X-ray to optical flux for type~1 AGN
(e.g., Figure~2 of Schmidt et~al. 1998), we should have detected
most type~1 AGN to $V\approx 23$ and might plausibly have detected 
some to $V\approx 27$. We also do not formally detect the 
$z=1.013$ strong FR~I radio source J123644+621133 (R98), although there is 
a hint of an X-ray photon excess at its position.   Among radio-loud AGN, however,
FR~I sources are the least X-ray luminous, and a rest-frame X-ray luminosity just
below our 2--10 keV upper limit of $\approx 2.2 \times 10^{42}$~erg~s$^{-1}$  would 
still make this source 
consistent with the 2--10 keV X-ray luminosities of other FR~I sources (compare with 
Table 3 of Sambruna, Eracleous \& Mushotzky 1999).
 
\subsection{Submillimeter Sources in the HDF-N and its Vicinity}

Deep submillimeter surveys have revealed a population of luminous, 
dusty, star-forming galaxies at moderate to high redshift that
make an important contribution to the energy output of the 
Universe (e.g., Hughes et~al. 1998, hereafter H98). However, it has
been difficult to determine the relative contributions that 
AGN and star formation make to powering these
objects (e.g., Almaini, Lawrence \& Boyle 1999). X-ray observations
may clearly discriminate between these possibilities.
 
H98 have presented 5 submillimeter sources in the HDF-N, and 
Barger et~al. (2000, hereafter B00) have presented an additional
5 sources in the vicinity of the HDF-N (here we consider submillimeter 
sources detected by B00 at $\geq 3\sigma$). All 10 of these 
sources are in the ACIS field, but we do not detect any of them. 
Two of these sources may show X-ray photon excesses near their 
positions, but further observations are needed before source 
detections can be claimed. 
We have used the \chandra\ data to place constraints on the 
primary energy generation mechanism for 
these 10 sources. We follow the general 
method described by Fabian et~al. (2000, hereafter F00), 
computing submillimeter (850~$\mu$m) to X-ray (2~keV) spectral 
indices ($\alpha_{\rm sx}$\footnote{Note that the `s' in $\alpha_{\rm sx}$ stands for `submillimeter' 
rather than `soft X-ray.'}) and comparing these with 
$\alpha_{\rm sx}$ values for local objects (see Table~2 and Figure~2). 
Because our \chandra\ observation is 8--18 times longer 
than those used by F00, we are able to place tight $\alpha_{\rm sx}$
constraints on a substantially larger number of submillimeter sources.

One straightforward interpretation of Figure~2 is that star 
formation is the primary energy source for these 10 submillimeter
sources.  Black hole accretion could only dominate the energy 
production if it is either (1) intrinsically X-ray weak or (2) obscured by 
Compton-thick material along the line of sight and there is little electron
scattering by a `mirror' in the nuclear environment. 
These are the tightest X-ray constraints yet on the submillimeter
source population, and the first to apply to enough 
sources to reasonably represent the population.
X-ray background synthesis models suggest 
that $\approx$~10--20\% of the submillimeter sources could 
be powered by AGN (Almaini et~al. 1999), and optical spectroscopy 
finds indications of AGN activity in at least $\approx 20$\% 
of submillimeter sources (Barger et~al. 1999). Our 0/10 rate
of X-ray detection is uncomfortably small, although it is
still statistically consistent with these values at the
$\sim 15$\% level. 
Although the redshifts used in Figure~2 have significant uncertainties 
(see H98 and B00), our basic results are not very sensitive to redshift. 

If we consider less-significant B00 
submillimeter sources in the 2.5--3$\sigma$ 
range, there are two additional $2.8\sigma$ submillimeter sources: 
123616.2+621513 (5.4~mJy, $z$ unknown) and 
123629.1+621046 (6.1~mJy, $z=1.013$). 
Both of these are positionally coincident with \chandra\ sources:
CXOHDFN~J123616.1+621513 (13.5 full-band counts) and 
CXOHDFN~J123629.1+621046 (23.6 full-band counts). 
The soft-band (hard-band) $\alpha_{\rm sx}$ values for 
these two sources are 
$>1.30$ ($=1.22$) and 
$=1.31$ ($>1.32$), respectively. 
These sources probably have some AGN activity, although such 
activity must be obscured or weak. 

\vspace*{0.10 in}

Additional \chandra\ observations during Cycles~1 and 2 will
allow an even deeper (by a factor of 3--6) X-ray survey of 
this area. Subsequent papers will discuss the X-ray sources detected 
in the full ACIS field of view and optical follow-up studies.


\acknowledgments

We thank
A. Barger,  
M.~Dickinson, 
M.~Eracleous,
A.~Fabian, 
D.~Hogg, 
K.~Iwasawa, 
C.~Liu, 
E.~Richards,  
C.~Steidel and 
R.~Williams for 
helpful discussions and kindly providing data.
We thank all the members of the \chandra\ team for their enormous efforts. 
We gratefully acknowledge the financial support of 
NASA grant NAS~8-38252 (GPG, PI), 
NASA GSRP grant NGT5-50247 (AEH), 
NASA LTSA grant NAG5-8107 and the Alfred P. Sloan Foundation (WNB), and  
NSF grant AST99-00703~(DPS). 



\begin{deluxetable}{lclcclccllll}
\rotate
\tablecolumns{11}
\tabletypesize{\scriptsize}
\tablenum{1}
\tablewidth{0pt}
\tablecaption {Properties of ACIS HDF-N Sources}
%
%
\tablehead{
\colhead{CXOHDFN Name}               & 
\colhead{Off axis}               &
\colhead{0.5--8.0 keV}   	 &     
\colhead{Band} 	       	         &
\colhead{$F_{0.5-8}$}            &
\colhead{$L_{0.5-8}$}            &
\colhead{W96}                    &
\colhead{CXO/W96} 	       	 &  
\colhead{} 	       	         &
\colhead{} 	       	         & 
\colhead{Descriptive} 	       	        
\\	
\colhead{(J2000)}                              &
\colhead{Angle ($^\prime$)}                    &
\colhead{Counts$^{\rm a}$}	               &	
\colhead{Ratio$^{\rm b}$} 		       & 
\colhead{(erg cm$^{-2}$ s$^{-1}$)$^{\rm c}$}   &
\colhead{(erg s$^{-1}$)$^{\rm d}$}             &	
\colhead{Name} 		                       &
\colhead{Offset ($^{\prime\prime}$)$^{\rm e}$} &
\colhead{Redshift$^{\rm f}$} 	               &
\colhead{$V_{606}^{\rm g}$} 	       	       & 
\colhead{Notes$^{\rm h}$} 	       	        
}
\startdata
123641.9+621131 & 2.65 &   $7.4\pm 2.8$  & $<1.92^{+2.10}_{-0.66}$ & $3.9\times 10^{-16}$ & $7.2\times 10^{39}$ & 4-976 & 1.92 & 0.089$^1$    & 20.0  & Spiral         \\
123646.4+621404 & 2.03 & $220.4\pm 15.0$ & $1.26^{+0.18}_{-0.16}$  & $1.2\times 10^{-14}$ & $3.3\times 10^{43}$ & 2-251 & 0.74 & 0.960$^2$    & 22.5  & AGN (R)        \\
123648.1+621309 & 1.51 &   $8.3\pm 3.0$  & $<1.02^{+0.67}_{-0.29}$ & $4.4\times 10^{-16}$ & $2.7\times 10^{41}$ & 2-121 & 0.62 & 0.476$^3$    & 21.4  & Elliptical     \\
123651.8+621220 & 1.23 &  $29.9\pm 5.6$  & $3.30^{+2.22}_{-1.16}$  & $3.1\times 10^{-15}$ & $1.7\times 10^{44}$ & ---   & 0.76 & 2.6--2.7$^4$ & 26.7  & NICMOS (R)     \\
123655.5+621310 & 0.68 &  $25.0\pm 5.1$  & $<0.36^{+0.11}_{-0.07}$ & $1.3\times 10^{-15}$ & $3.9\times 10^{42}$ & 3-180 & 0.73 & 0.968$^5$    & 23.2  & Elliptical (R) \\
123657.0+621301 & 0.48 &  $10.2\pm 3.3$  & $<1.16^{+0.86}_{-0.35}$ & $5.4\times 10^{-16}$ & $3.2\times 10^{41}$ & 3-355 & 0.53 & 0.474$^6$    & 24.4  & Elliptical (R) \\
\enddata
\tablenotetext{a}{Source counts and errors are as computed by
{\sc wavdetect} (see Freeman et~al. 2000 for details).}
\tablenotetext{b}{Defined as the ratio of counts between the 2--8~keV
and 0.5--2~keV bands. Errors for this quantity are calculated following
the `numerical method' described in \S1.7.3 of Lyons (1991).}
\tablenotetext{c}{Fluxes in this column are for the observed
frame and are not corrected for Galactic 
absorption. For all sources other than 123651.8+621220, 
fluxes have been computed adopting a power-law model with photon 
index $\Gamma=2$ and the Galactic column density. For 123651.8+621220
we also include a column density of $10^{23}$~cm$^{-2}$ at 
$z=2.6$ (see \S3.1).}
\tablenotetext{d}{Luminosities are for the rest frame and are
corrected for Galactic absorption. We again adopt a power-law
model with $\Gamma=2$. 
For 123651.8+621220 we also correct for the probable intrinsic 
absorption (see note~c), and we adopt the photometric redshift
estimate of $z=2.6$ for lack of a spectroscopic redshift 
(see \S3.1). The 2--8~keV luminosity for 123651.8+621220 
is $9.3\times 10^{43}$~erg~s$^{-1}$, and this quantity is 
relatively insensitive to corrections for internal absorption.}
\tablenotetext{e}{Offset between the \chandra\ source and the proposed
W96 identification. For 123651.8+621220 we quote the offset from
NICMOS~J123651.74+621221.4.}
\tablenotetext{f}{Redshift references are indicated by numerical superscripts as 
follows: 
(1) Cohen et~al. (1996), 
(2) Phillips et~al. (1997), 
(3) Hogg et~al. (2000), 
(4) M. Dickinson, private communication, 
(5) Hogg et~al. (2000), 
(6) Barger et~al. (2000).
The redshift for CXOHDFN~123651.8+621220 is photometric and subject to
significant uncertainty (see \S3.1).}
\tablenotetext{g}{All $V_{606}$ magnitudes other than that for 
123651.8+621220 are from W96. The $V_{606}$ magnitude for
123651.8+621220 is from M. Dickinson, private communication.}
\tablenotetext{h}{Sources with `(R)' are 8.4~GHz sources from R98.}
\end{deluxetable}

\clearpage


\begin{deluxetable}{lccclllll}
\tabletypesize{\footnotesize}
\tablenum{2}
\tablewidth{0pt}
\tablecaption {\chandra\ Constraints on 850~$\mu$m Sources}
%
\scriptsize
\tablehead{
\colhead{Source}                               & 
\colhead{$S_{850}$}                            &
\colhead{}                                     &     
\colhead{0.5--2~keV $\alpha_{\rm sx}$/}        \\	
\colhead{Name}                               &
\colhead{(mJy)$^{\rm a}$}                    &
\colhead{$F_{0.5-2}$/$F_{2-8}^{\rm b}$}      &	
\colhead{2--8~keV $\alpha_{\rm sx}^{\rm c}$} &
}
\startdata
HDF~850.1        &  7.0 & $<1.5$ / $<8.3$     & $>1.37$ / $>1.26$  & \\   
HDF~850.2        &  3.8 & $<2.1$ / $<5.8$     & $>1.31$ / $>1.24$  & \\
HDF~850.3        &  3.0 & $<1.5$ / $<8.3$     & $>1.32$ / $>1.20$  & \\
HDF~850.4        &  2.3 & $<2.1$ / $<10.5$    & $>1.28$ / $>1.17$  & \\
HDF~850.5        &  2.1 & $<1.5$ / $<5.8$     & $>1.29$ / $>1.20$  & \\
123618.3+621551  &  7.8 & $<1.6$ / $<9.0$     & $>1.38$ / $>1.26$  & \\  
123621.3+621708  &  7.5 & $<2.3$ / $<6.5$     & $>1.36$ / $>1.28$  & \\  
123622.7+621630  &  7.1 & $<2.2$ / $<9.0$     & $>1.35$ / $>1.26$  & \\  
123646.1+621449  & 10.7 & $<2.7$ / $<13.2$    & $>1.37$ / $>1.26$  & \\  
123700.3+620910  & 11.9 & $<1.6$ / $<9.0$     & $>1.41$ / $>1.29$  & \\  
%
%
\enddata
\tablenotetext{a}{850~$\mu$m flux density.}
\tablenotetext{b}{Fluxes in this column are in units of 
$10^{-16}$~erg~cm$^{-2}$~s$^{-1}$. They are for the observed 
frame and are not corrected for Galactic absorption. They
have been computed adopting a power-law model with photon 
index $\Gamma=2$ and the Galactic column density. 
Upper limits are calculated using the Bayesian method 
of Kraft, Burrows \& Nousek (1991) for 99\% confidence;
the uniform prior used by these authors results in fairly 
conservative upper limits, and other reasonable choices of priors 
do not materially change our scientific results.}
\tablenotetext{c}{The `0.5--2~keV $\alpha_{\rm sx}$' uses a 2~keV flux
density derived by converting the observed 0.5--2~keV counts into a
flux. The `2--8~keV $\alpha_{\rm sx}$' is derived in a similar manner
using the 2--8~keV counts. Spectral models are as per note~b, and the
flux densities used are for the observed frame.}  
\end{deluxetable}

\clearpage


\begin{figure}
\rotate
\epsscale{0.8}
\plotone{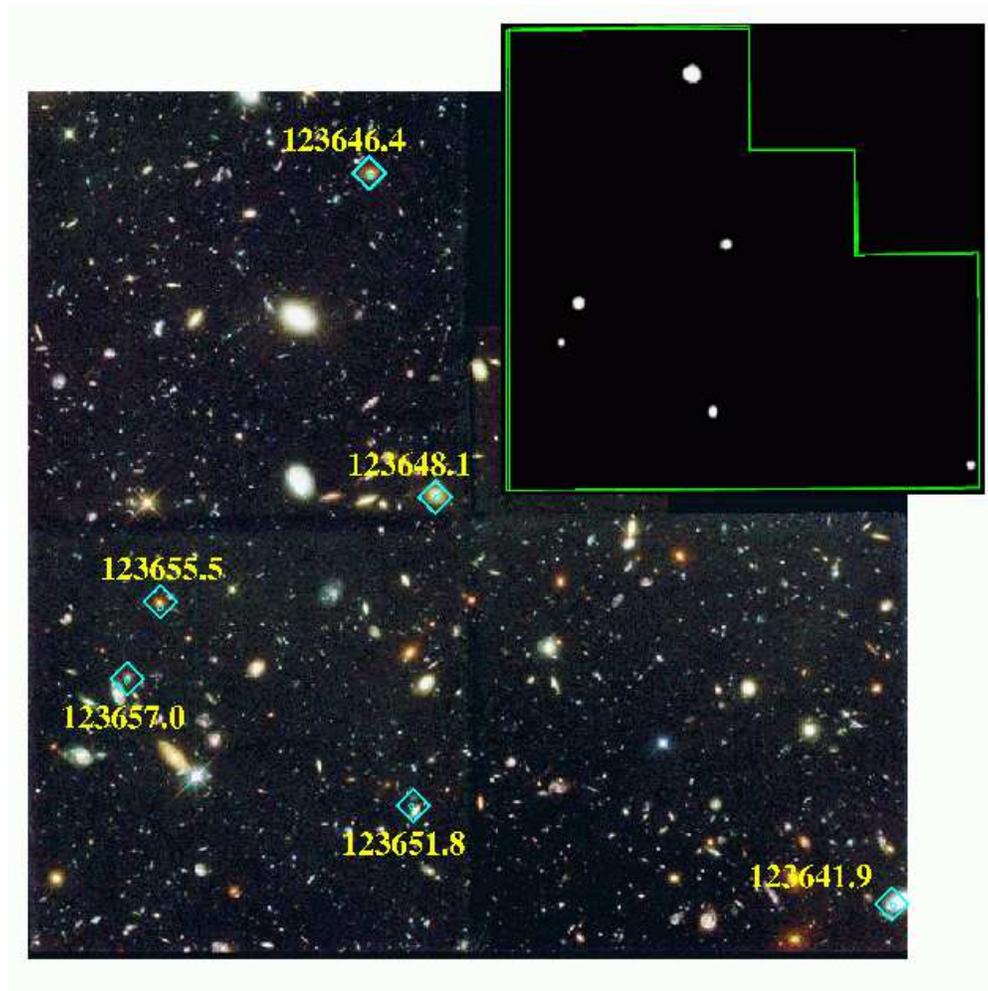}
\vspace{0.5in}
\caption{The main image shows the \chandra\ sources detected in the HDF-N
overlaid on the W96 optical image. \chandra\ sources are labeled by the right 
ascension parts of their CXO names (see Table~1). The small ellipses inside the 
diamonds have sizes that approximately match our statistical positional 
uncertainties (the diamonds are drawn only to help the reader locate
the ellipses). The smaller image to the upper right shows the \chandra\ image
after adaptive smoothing using the code of Ebeling, White \& Rangarajan (2000). 
The smoothing has been performed at the $2.5\sigma$ level.  
\label{fig1}}
\end{figure}
 


\begin{figure}
\epsscale{0.8}
\plotone{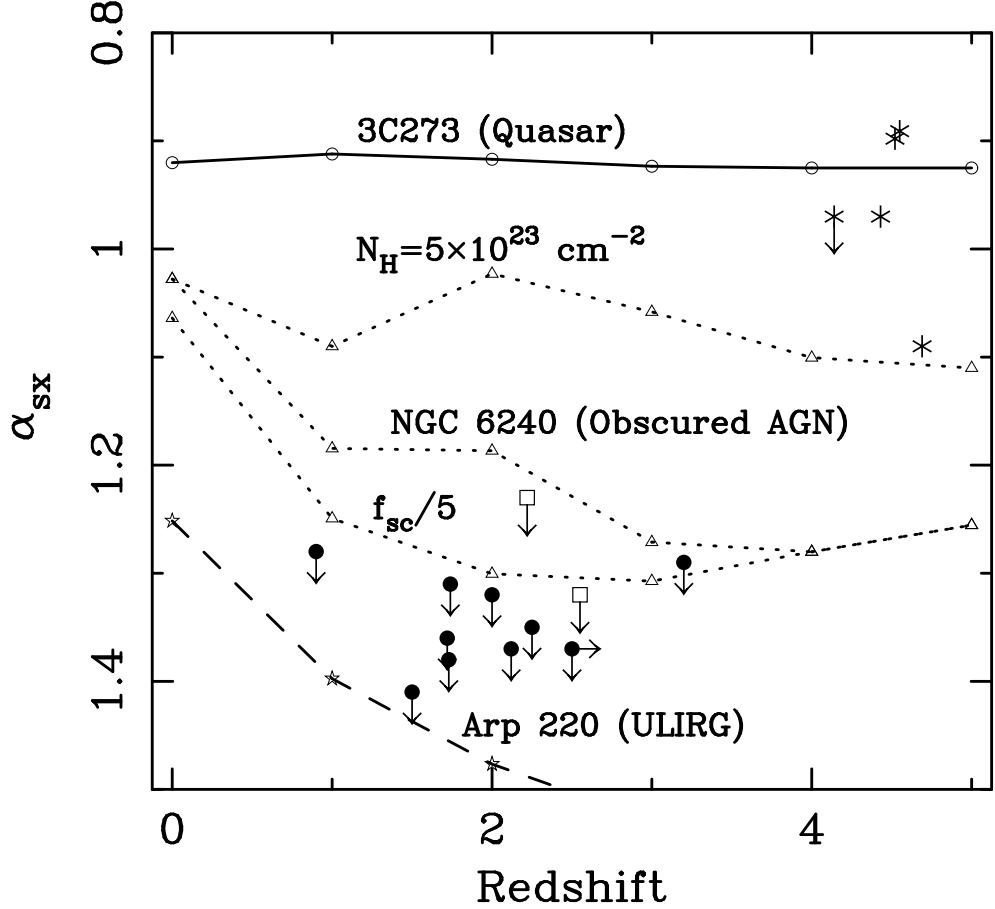}
\vspace*{0.5in}
\caption{Submillimeter to X-ray spectral index ($\alpha_{\rm sx}$)
plotted against redshift (adapted from F00). 
Shown are 
the quasar 3C273 (solid curve), 
the ultraluminous obscured 
($N_{\rm H}\approx 2\times 10^{24}$~cm$^{-2}$) AGN NGC~6240 (dotted curves), 
the ultraluminous infrared galaxy (ULIRG) Arp~220 (dashed curve), 
five $z>4$ quasars (asterisks), 
two submillimeter sources from F00 (open squares), and 
the 10 submillimeter sources from the HDF-N and its vicinity (solid dots; these are
our 0.5--2~keV $\alpha_{\rm sx}$ values).
For NGC~6240, the effects of decreasing its column density to
$5\times 10^{23}$~cm$^{-2}$ and its scattered X-ray fraction 
by a factor of 5 are also shown. 
For the HDF-N area submillimeter sources we have calculated 
millimetric redshifts following B00 when possible; see B00 for a 
discussion of the errors associated with these. For HDF~850.3, 
HDF~850.4 and HDF~850.5 we have used the redshift estimates from 
Table~2 of H98. 
\label{fig2}}
\end{figure}
 


\end{document}